\UseRawInputEncoding
\documentclass[runningheads]{llncs}
\usepackage{graphicx}
\usepackage{blindtext} 
\usepackage{url}
\usepackage{xurl}
\usepackage{ amssymb }
\usepackage{amsmath}

\begin{document}
\title{Transformers for CT Reconstruction From Monoplanar and Biplanar Radiographs}
%
%
\author{Firas Khader\inst{1} \and
Gustav M\"uller-Franzes\inst{1} \and
Tianyu Han\inst{2} \and
Sven Nebelung\inst{1} \and
Christiane Kuhl\inst{1} \and
Johannes Stegmaier\inst{3} \and
Daniel Truhn\inst{1}
}
%
\authorrunning{F. Khader et al.}
%
\institute{
 Department of Diagnostic and Interventional Radiology, University Hospital Aachen, Aachen, Germany \and
Physics of Molecular Imaging Systems, Experimental Molecular Imaging, RWTH Aachen University, Aachen, Germany \and
Institute of Imaging and Computer Vision, RWTH Aachen University, Aachen, Germany
}

\maketitle              
\begin{abstract}
Computed Tomography (CT) scans provide detailed and accurate information of internal structures in the body. They are constructed by sending x-rays through the body from different directions and combining this information into a three-dimensional volume. Such volumes can then be used to diagnose a wide range of conditions and allow for volumetric measurements of organs. In this work, we tackle the problem of reconstructing CT images from biplanar x-rays only. X-rays are widely available and even if the CT reconstructed from these radiographs is not a replacement of a complete CT in the diagnostic setting, it might serve to spare the patients from radiation where a CT is only acquired for rough measurements such as determining organ size. We propose a novel method based on the transformer architecture, by framing the underlying task as a language translation problem. Radiographs and CT images are first embedded into latent quantized codebook vectors using two different autoencoder networks. We then train a GPT model, to reconstruct the codebook vectors of the CT image, conditioned on the codebook vectors of the x-rays and show that this approach leads to realistic looking images. To encourage further research in this direction, we make our code publicly available on GitHub: XXX.

\keywords{Chest Radiography \and Computed Tomography \and Multi Modal \and Transformers.}
\end{abstract}
\section{Introduction}
In clinical practice, two widely adopted imaging techniques are X-ray radiographs and computed tomography (CT) scans. They serve as essential diagnostic tools for radiologists and are used to examine the human body as well as detect abnormalities or injuries \cite{khader_artificial_2022,dey_diagnostic_2018} While both employ X-rays to provide radiologists with an image of the patients body, radiographs are only capable of generating two-dimensional (2D) projections of the body, while CT scans allow for more detailed three-dimensional (3D) image. The latter is possible, as CT images are computationally constructed by composing multiple X-rays taken from different directions into a detailed cross-section image of the body. As a longer exposure time is required to construct such 3D volumes, CT scans are associated with a high radiation dose. However, in some cases, a detailed image is not necessary, and CT imaging is used solely to measure the size and extent of organs like the liver before surgery. Radiographs, on the other hand, cannot provide quantitative measurements in terms of volume, but they are more cost-effective and radiation-friendly than CT scans. 
\newline\newline
As both imaging modalities are based on X-rays, methods exist to transform the 3D CT scans into matching 2D radiographs. However, this transformation leads to a loss of information, as many details of the image will be removed. Thus, the opposite direction, i.e., transforming a 2D radiograph into a 3D CT scan is not easily possible. Nonetheless, enabling such reconstructions can provide significant benefits, as synthesizing a CT image from existing radiographs could help reduce the patient's exposure to radiation from a CT scan and assist surgeons in their planning process, allowing them to more accurately assess the patient's anatomy.
\newline\newline
Over the years, several methods have been proposed that attempt to convert radiographs into corresponding CT scans \cite{ying_x2ct-gan_2019-1,shiode_2d3d_2021,shen_patient-specific_2019,schock_monoplanar_2022}. These methods are largely based on convolutional neural networks (CNNs) and employ GAN-based \cite{goodfellow_generative_2014} discriminators to synthesize realistic looking images. However, such network designs typically require unsymmetric encoder-decoder networks, in which a separate encoder branch is needed for each 2D projection \cite{ying_x2ct-gan_2019-1}. Moreover, GAN-based methods are known to suffer from mode collapse, resulting in similar looking images \cite{arjovsky_wasserstein_2017}.
\newline\newline
More recently, the transformer architecture \cite{vaswani_attention_2017} has been introduced and has shown a remarkable performance in text-to-image generation tasks \cite{ramesh_zero-shot_2021,esser_taming_2021}. New images are thereby generated in an autoregressive manner by modelling the joint distribution over the input tokens. Contrary to previous methods that rely on CNNs, in this study we built upon the recent progress of the transformer architecture and frame the cross-modality transfer as a language translation problem and by doing so leverage the power of transformers to convert monoplanar and biplanar radiographs into corresponding CT scans.

\section{Materials \& Methods}
\begin{figure}[h]
\centering
\includegraphics[width=0.76\textwidth]{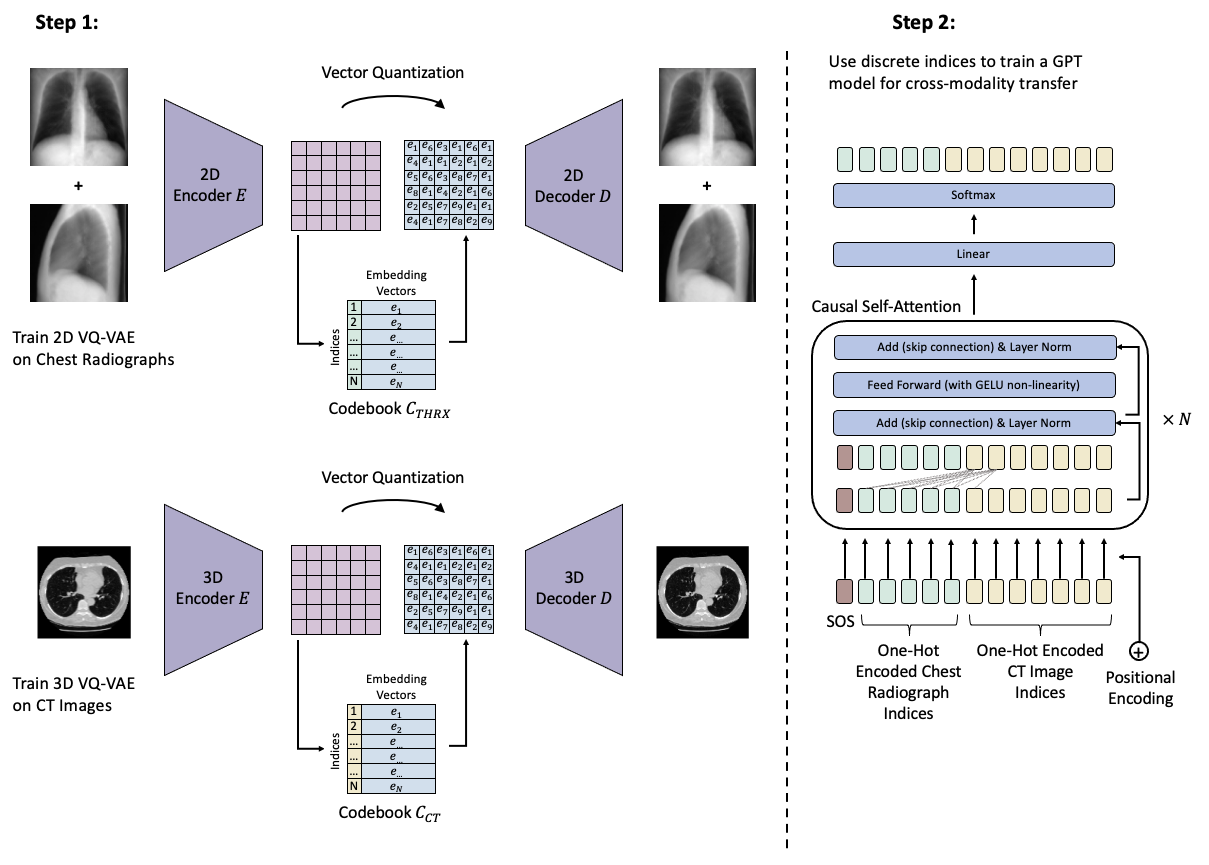}
\caption{Model architecture. In a first step, two different VQ-GAN models are trained: A 2D version which is used to transform the biplanar chest radiographs into a set of discrete codebook indices and a 3D version used to convert the CT image volumes into another set of codebook indices. By framing the underlying problem as a language translation problem, we then train a GPT model that, given the indices of the chest radiographs, autoregressively learns to generate indices corresponding to the CT volumes.} 
\label{training}
\end{figure}

\subsection{Dataset}
We train and test our model on the publicly available LIDC-IDRI dataset \cite{armato_iii_data_2015,armato_lung_2011,clark_cancer_2013} (\url{https://wiki.cancerimagingarchive.net/}). This dataset consists of 1,018 diagnostic and lung-cancer screening thoracic CT scans from 1,010 patients. The data was collected from seven academic centers and eight medical imaging companies. In order to use this data to train our models, we apply a number of pre-processing steps: First, we convert the voxel values of each CT into Hounsfield units. Subsequently, we resample each volume into an isotropic voxel spacing of 1mm in all directions and center crop or pad each volume such that the resulting image is of shape $320 \times 320 \times 320$ (height, width, depth). In a final step, we resize the image into the shape $120 \times 120 \times 120$ and normalize each image to the range between -1 and 1. We proceed by splitting the pre-processed data into a training (70\%, n=707 patients), validation (20\%, n=202 patients) and testing split (20\%, n=101 patients).
\subsubsection{Digitally Reconstructed Radiographs}
\begin{figure}[h]
\centering
\includegraphics[width=0.57\textwidth]{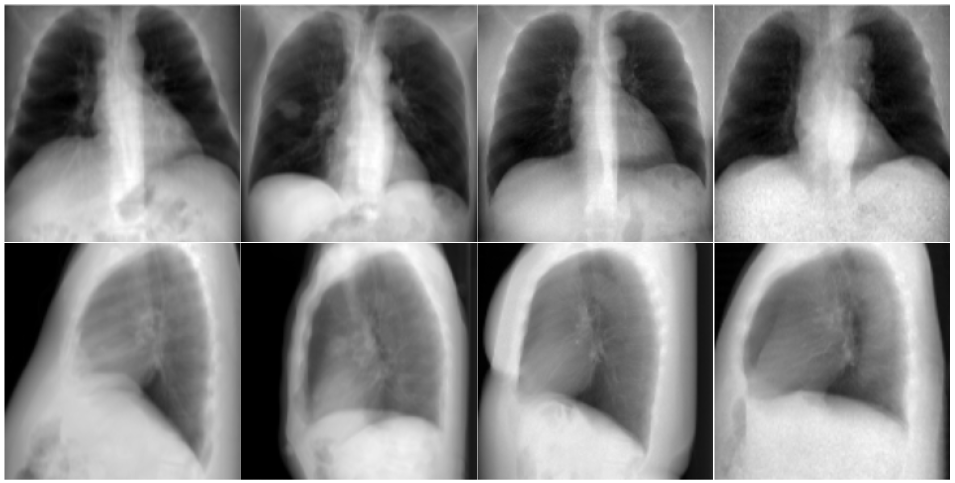}
\caption{Digitally reconstructed radiographs. To create a set of corresponding chest radiographs and CT images, we project each CT image into a posterior-anterior view radiograph and a lateral view radiograph.} \label{radiographs}
\end{figure}
The dataset mentioned above consists primarily of CT images without corresponding radiographs captured in lateral or posterior-anterior view. Consequently, we generated synthetic chest radiographs in both views (see Figure \ref{radiographs}) via digitally reconstructed radiographs \cite{milickovic_ct_2000}. More precisely, given a CT volume $x$ and the voxel depth $d_i$ (in direction of the projection) for each voxel $i$ in the image, we first convert the voxel values from Hounsfield units to their corresponding linear attenuation coefficients $u_i$. The digitally reconstructed radiograph can then be constructed by computing the projection given by 
\begin{equation}
   x_\text{projection} = I_0 \exp(- \sum_{i=1}^{n}{u_i d_i})
\end{equation}
The factor $I_0$ thereby denotes the signal intensity of the X-ray photons that we arbitrarily set to 1keV and $n$ denotes the number of voxels that the ray passes through.

\subsection{Model}
In contrast to previous work that mainly proposes the use of unsymmetric CNN-based encoder-decoder architectures with a discriminator loss for converting radiographs into CT scans, we make use of a transformer-based method (see Figure \ref{training}). Fundamentally, we frame the underlying task as a language translation problem, in which a set of discrete tokens from language A (i.e., the radiographs) is translated into a set of discrete tokens from language B (i.e., the CT scans). In the following we will briefly introduce the concept of vector quantization with regard to the VQ-VAE \cite{van_den_oord_neural_2017} and the VQ-GAN \cite{esser_taming_2021} models. Subsequently, we show how the GPT \cite{brown_language_2020} model can then be used to perform the modality transfer.

\subsubsection{Vector Quantization}
In order to treat the underlying problem as a language translation problem, where we go from a discrete set of tokens from language A to a discrete set of tokens from language B, we first have convert the two-dimensional chest radiographs and the three-dimensional CT volumes into such discrete representations. One popular approach for obtaining discrete representations with a reasonably sized alphabet (represented as a codebook in the following) is through the use of a VQ-VAE \cite{van_den_oord_neural_2017}. In essence, an encoder neural network $E$ is used to compress the image into a latent dimension with a continuous feature space. This feature space is then quantized by mapping each feature vector onto its nearest neighbor in a learnable codebook $C$ with $N$ entries. The resulting quantized feature representation is fed into a decoder neural network $D$ trained to reconstruct the image. The loss can therefore be expressed as
\begin{equation}
    \mathcal{L_{VQ}} = \|x - \hat{x}\|^2 + \|\text{sg}[E(x)] - e \|^2_2 + \|\text{sg}[e] - E(x) \|^2_2
\end{equation}
where $x$ denotes the input, $\hat{x}$ denotes the reconstructed output, $e$ represents the vector quantized feature vector and $\text{sg[.]}$ indicates the stop-gradient operation. Following Esser et al.'s improvements in their proposed VQ-GAN model \cite{esser_taming_2021}, we add a perceptual loss  \cite{johnson_perceptual_2016} $\mathcal{L}_P$ and a discriminator loss \cite{isola_image--image_2017-1} $\mathcal{L_G}$ to the output of the decoder. The discriminator is thereby trained to distinguish between the real input image $x$ and the reconstructed output image $\hat{x}$, which has been found to produce superior reconstructions. We extend the original 2D formulation of the model to also support 3D images by replacing the 2D convolution operations by 3D operations and modifying the perceptual loss term to be the mean perceptual loss over all depth slices in the 3D volume. Given both versions of the model, we train the 2D VQ-GAN to create a discrete representation for the lateral and posterior-anterior views of the radiographs (i.e., one network is trained for both views) with a codebook $C_\text{THRX}$ of size $N$. Similarly, we train the 3D VQ-GAN on the CT images using a codebook $C_{\text{CT}}$ of the same size $N$.

\subsubsection{GPT-based Language Translation}
To perform the task of translating token representations of the biplanar chest radiographs, which are specified by indices in the codebook $C_{\text{THRX}}$, into token representations associated with their corresponding CT image, which are specified by indices in the codebook $C_{\text{CT}}$, we train an autoregressive GPT model. More precisely, we first convert the set of indices that represent each image into a one-hot encoded representation (i.e., the codebook vector at position 3 would be encoded into the vector representation $(0, 0, 1, ...)^T$ of size $N$). Subsequently we order the resulting token representations by first listing all tokens pertaining to the radiograph in posterior-anterior view, followed by the tokens pertaining to the radiograph in lateral view and finally the token representations of the ground truth CT scan. Additionally, a start of sentence token (SOS) is prepended to the set of tokens, which is used to later prompt the GPT model with the token generation. A learnable positional encoding vector is then added to each token. These token representations are then sequentially passed through a series of transformer blocks, where the number of blocks $l$ is set to 8 in our model. Within each transformer block, a causal self-attention mechanism is employed, thereby restricting each token from attending to tokens that come after it in the sequence, allowing it to only attend to tokens that precede it. The output of the last transformer block is subsequently passed through a linear layer as well as a softmax layer to arrive at the prediction, which represents the probability of the next token in the sequence. The cross-entropy loss is used as the loss function.

\subsubsection{Inference}
\begin{figure}[h]
\centering
\includegraphics[width=0.76\textwidth]{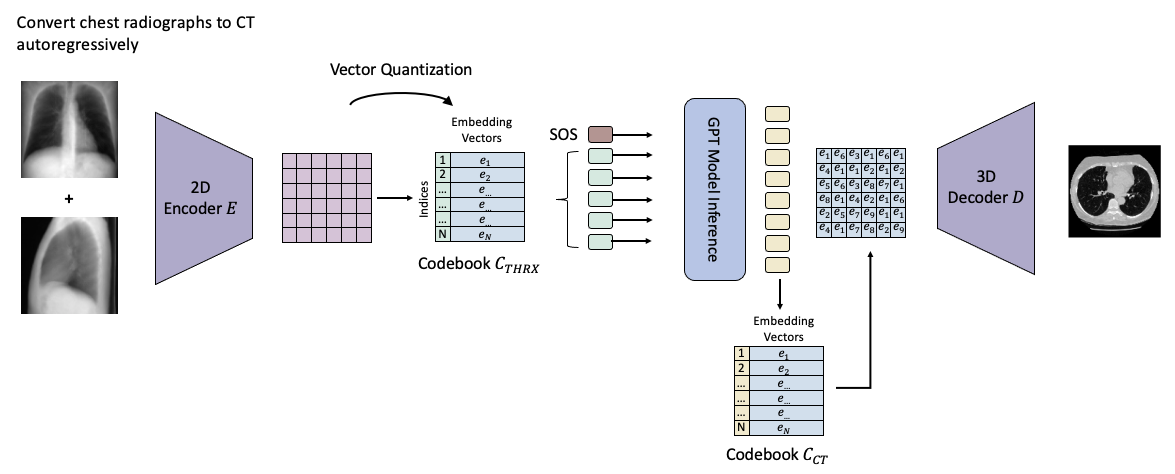}
\caption{Inference procedure. After the respective models have been trained we can convert a set of unseen chest radiographs to a synthetic CT volume by first converting the chest radiographs into their discrete latent representation. The codebook indices of this latent representation are then used to autoregressively infer new set of indices that correspond to the latent representation of the CT volume. In a final step, the new set of discrete indices is converted to the CT volume by feeding them into the decoder of the previously trained 3D VQ-GAN} \label{inference}
\end{figure}
After completing the training process for all networks, we can proceed to generate synthetic CT images. To achieve this, we first transform a set of unseen radiographs captured in posterior-anterior and lateral views into their discrete token representations using the previously trained 2D VQ-GAN. These representations are then concatenated, such that the tokens originating from the posterior-anterior view are listed first, followed by the tokens from the lateral view. Subsequently, we feed this concatenated representation into the trained GPT model to prompt the generation of tokens that correspond to the CT images in an autoregressive manner. The generation process continues until a total of $J$ tokens are generated, where $J$ denotes the number of indices required to represent the latent representation of the CT image. In a final step, we utilize the previously trained decoder of the 3D VQ GAN to convert the latent representation of the CT image into an actual synthetic CT image (see Figure \ref{inference}). It is worth noting that this method also enables the creation of synthetic CT images when using solely a single radiograph as input (i.e., the radiograph captured in posterior-anterior view). This is achieved by feeding the single radiograph to the GPT model and prompting it to generate tokens that correspond to the radiograph captured in the lateral view and the CT image.

\subsection{Training Settings}
All our models were trained on an NVIDIA RTX A6000 and were implemented using PyTorch v2.1.0. The VQ-GAN models were trained for 500 epochs and we chose the models that performed best (i.e., resulted in the lowest loss) on the validation set to encode and decode the images. The GPT model was trained for a total of 300 epochs, and similarly, the model that demonstrated the best performance in terms of the lowest loss on the validation set was selected for evaluation. For more details regarding the used hyperparameters, please refer to Supplementary Table \ref{tab:hyperparams}

\section{Results}
\begin{figure}[h]
\centering
\includegraphics[width=0.7\textwidth]{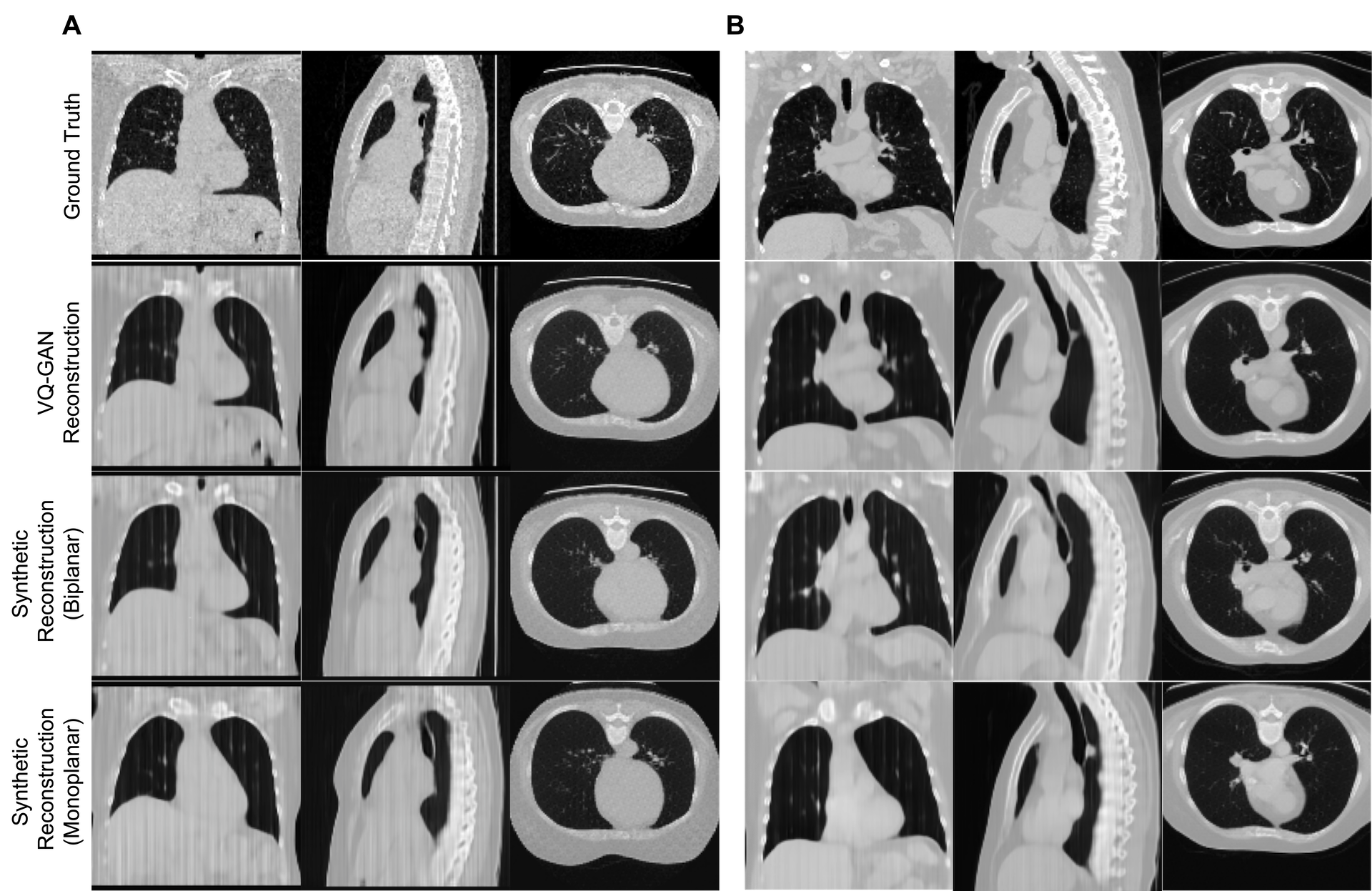}
\caption{Synthetic CT images produced by our model using radiographs from the test dataset. (A) and (B) denote two different samples. The top row shows the ground truth CT images, while the second row shows reconstructed CT images using the trained 3D VQ GAN model. The third row displays synthetic CT images generated by our model when presented with radiographs taken in posterior-anterior and lateral views, and the fourth row displays synthetic CT images generated by the model when only provided with a single radiograph in posterior-anterior view.} \label{combined sample}
\end{figure}
\subsection{Biplanar CT Reconstruction}
For biplanar reconstruction, i.e. the reconstruction of the full CT volume from both a posterior-anterior and a lateral radiograph, we found that the generated volume looked realistic as a whole, but that fine details of the organs such as lung structure or the internal structure of the heart or the bones were not reconstructed faithfully. 
A radiologist with 10 years of experience compared the GPT-reconstructed CT volumes (see Figure \ref{combined sample}) using radiographs from the test dataset with the original volumes and rated the following categories on a scale from 0 (= no agreement between original and reconstructed structure) to 5 (= perfect agreement between original and reconstructed structure): Inner structure of the heart (distribution of heart chambers and vessels): 1.8 $\pm$ 1.2.
Form of the heart (outline of the heart shape):  4.2 $\pm$ 0.7.
Inner structure of the lungs (distribution of vessels and bronchi): 1.7 $\pm$ 1.5.
Form of the lungs (outline of the lungs): 4.3 $\pm$ 0.6. Inner structure of bones (can structure be inferred from the bones, such as osteoporotic changes?): 0.7 $\pm$ 0.3.
General outline of bone (does the vertebral spine follow the same pattern, e.g. skoliosis, kyphosis): 4.1 $\pm$ 0.7.
These categories were chosen to both rate fine details (inner structure) and the more general anatomical outline of larger structures. The results were in line with expectations as organ outlines can already be seen on radiographs, while the inner structures are barely visible on radiographs due to the superposition effect.

\subsection{Monoplanar CT Reconstruction}
In the clinical setting, radiographs are not routinely performed in both posterior-anterior and lateral acquisition schemes, but often only in posterior-anterior position. Previous methods either involve modifying the encoder branches to accommodate the varying number of input images or utilize PCA or k-nearest neighbor methods to generate the missing views \cite{schock_monoplanar_2022}. In contrast, our architecture is flexible in the sense that it can accomodate for missing data and reconstruct full CT volumes from posterior-anterior radiographs alone without any changes to the model. Therefore, we generated CT examinations from these radiographs only and repeated the experiments with the radiologist. The ratings were slightly decreased, but showed the same trend:
Inner structure of the heart: 1.7 $\pm$ 1.1.
Form of the heart:  3.9 $\pm$ 0.8.
Inner structure of the lungs: 1.7 $\pm$ 1.6.
Form of the lungs: 3.9 $\pm$ 0.7.
Inner structure of bones: 0.7 $\pm$ 0.4.
General outline of bone: 4.1 $\pm$ 0.6.
Again, this was in line with our expectations: As before, it can be seen that organ outlines are easier to reconstruct than fine details and it can also be expected that the reconstruction is less faithful as compared to the situation where an additional lateral radiograph serves as an additional source of information.

\section{Conclusion}
In this work, we developed and presented a novel transformer-based method capable of converting a set of 2D radiographs into a corresponding 3D CT scan. We demonstrate that this model can generate CT scans that largely capture the outline and form of the heart, lungs and bones accurately. This information can potentially be used clinically to measure the size of organs, as for such purposes only the general outline is necessary. We also found that fine details, such as the parenchymal structure of the lung, or the internal structure of the heart can not reliably be reconstructed. This is understandable since this information is most likely not sufficiently presented by two projections alone. Importantly, this is a limitation that is shared by all methods of creating 3D volumes from 2D radiographs, but that is often overlooked and not sufficiently evaluated.
Furthermore, we show that the transformer-based method allows for a reconstruction of CT images based on single posterio-anterior radiographs, without the need of explicitly modifying any parts of the architecture.
In addition, incorporating information of other modalities, such as laboratory values which are obtained next to the radiographs, can easily be performed by inputting these elements as tokens into the GPT model. This could be useful in cases where these laboratory values are informative of structural changes (e.g., brain natriuretic peptide as an indicator for enlargement of certain heart chambers). In future work we will assess the performance of our model on large-scale CT image databases to fully leverage the power of scaleable transformer architectures.

\section{Acknowledgements}
The authors acknowledge the National Cancer Institute and the Foundation for the National Institutes of Health, and their critical role in the creation of the free publicly available LIDC/IDRI Database used in this study.

%
%
%
\newpage
\bibliographystyle{splncs04}
\bibliography{references_cleaned}

\begin{thebibliography}{10}
\providecommand{\url}[1]{\texttt{#1}}
\providecommand{\urlprefix}{URL }
\providecommand{\doi}[1]{https://doi.org/#1}

\bibitem{arjovsky_wasserstein_2017}
Arjovsky, M., Chintala, S., Bottou, L.: Wasserstein {GAN} (Dec 2017),
  \url{http://arxiv.org/abs/1701.07875}, arXiv:1701.07875 [cs, stat]

\bibitem{armato_lung_2011}
Armato, S.G., et~al.: The {Lung} {Image} {Database} {Consortium} ({LIDC}) and
  {Image} {Database} {Resource} {Initiative} ({IDRI}): {A} {Completed}
  {Reference} {Database} of {Lung} {Nodules} on {CT} {Scans}. Medical Physics
  \textbf{38}(2),  915--931 (Feb 2011),
  \url{https://www.ncbi.nlm.nih.gov/pmc/articles/PMC3041807/}

\bibitem{armato_iii_data_2015}
Armato~III, S.G.: Data {From} {LIDC}-{IDRI} (2015),
  \url{https://wiki.cancerimagingarchive.net/x/rgAe}, version Number: 4 Type:
  dataset

\bibitem{brown_language_2020}
Brown, T.B., Mann, B., Ryder, N., Subbiah, M., Kaplan, J., Dhariwal, P.,
  Neelakantan, A., Shyam, P., Sastry, G., Askell, A., Agarwal, S.,
  Herbert-Voss, A., Krueger, G., Henighan, T., Child, R., Ramesh, A., Ziegler,
  D.M., Wu, J., Winter, C., Hesse, C., Chen, M., Sigler, E., Litwin, M., Gray,
  S., Chess, B., Clark, J., Berner, C., McCandlish, S., Radford, A., Sutskever,
  I., Amodei, D.: Language {Models} are {Few}-{Shot} {Learners} (Jul 2020),
  \url{http://arxiv.org/abs/2005.14165}, arXiv:2005.14165 [cs]

\bibitem{clark_cancer_2013}
Clark, K., Vendt, B., Smith, K., Freymann, J., Kirby, J., Koppel, P., Moore,
  S., Phillips, S., Maffitt, D., Pringle, M., Tarbox, L., Prior, F.: The
  {Cancer} {Imaging} {Archive} ({TCIA}): {Maintaining} and {Operating} a
  {Public} {Information} {Repository}. Journal of Digital Imaging
  \textbf{26}(6),  1045--1057 (Dec 2013),
  \url{https://www.ncbi.nlm.nih.gov/pmc/articles/PMC3824915/}

\bibitem{dey_diagnostic_2018}
Dey, R., Lu, Z., Hong, Y.: Diagnostic classification of lung nodules using {3D}
  neural networks. In: 2018 {IEEE} 15th {International} {Symposium} on
  {Biomedical} {Imaging} ({ISBI} 2018). pp. 774--778 (Apr 2018), iSSN:
  1945-8452

\bibitem{esser_taming_2021}
Esser, P., Rombach, R., Ommer, B.: Taming {Transformers} for
  {High}-{Resolution} {Image} {Synthesis}. In: Proceedings of the {IEEE}/{CVF}
  {Conference} on {Computer} {Vision} and {Pattern} {Recognition}. pp.
  12873--12883 (2021),
  \url{https://openaccess.thecvf.com/content/CVPR2021/html/Esser_Taming_Transformers_for_High-Resolution_Image_Synthesis_CVPR_2021_paper.html}

\bibitem{goodfellow_generative_2014}
Goodfellow, I., Pouget-Abadie, J., Mirza, M., Xu, B., Warde-Farley, D., Ozair,
  S., Courville, A., Bengio, Y.: Generative {Adversarial} {Nets}. In: Advances
  in {Neural} {Information} {Processing} {Systems}. vol.~27. Curran Associates,
  Inc. (2014)

\bibitem{isola_image--image_2017-1}
Isola, P., Zhu, J.Y., Zhou, T., Efros, A.A.: Image-to-{Image} {Translation}
  with {Conditional} {Adversarial} {Networks}. In: 2017 {IEEE} {Conference} on
  {Computer} {Vision} and {Pattern} {Recognition} ({CVPR}). pp. 5967--5976 (Jul
  2017). \doi{10.1109/CVPR.2017.632}, iSSN: 1063-6919

\bibitem{johnson_perceptual_2016}
Johnson, J., Alahi, A., Fei-Fei, L.: Perceptual {Losses} for {Real}-{Time}
  {Style} {Transfer} and {Super}-{Resolution}. In: Computer {Vision} – {ECCV}
  2016. pp. 694--711. Lecture {Notes} in {Computer} {Science}, Springer
  International Publishing, Cham (2016)

\bibitem{khader_artificial_2022}
Khader, F., Han, T., Müller-Franzes, G., Huck, L., Schad, P., Keil, S.,
  Barzakova, E., Schulze-Hagen, M., Pedersoli, F., Schulz, V., Zimmermann, M.,
  Nebelung, L., Kather, J., Hamesch, K., Haarburger, C., Marx, G., Stegmaier,
  J., Kuhl, C., Bruners, P., Nebelung, S., Truhn, D.: Artificial {Intelligence}
  for {Clinical} {Interpretation} of {Bedside} {Chest} {Radiographs}. Radiology
  p. 220510 (Dec 2022), \url{https://pubs.rsna.org/doi/10.1148/radiol.220510},
  publisher: Radiological Society of North America

\bibitem{kingma_adam_2015-1}
Kingma, D.P.: Adam: {A} {Method} for {Stochastic} {Optimization}.  (2015),
  \url{http://arxiv.org/abs/1412.6980}

\bibitem{loshchilov_sgdr_2017}
Loshchilov, I., Hutter, F.: {SGDR}: {Stochastic} {Gradient} {Descent} with
  {Warm} {Restarts} (May 2017), \url{http://arxiv.org/abs/1608.03983},
  arXiv:1608.03983 [cs, math]

\bibitem{loshchilov_decoupled_2019}
Loshchilov, I., Hutter, F.: Decoupled {Weight} {Decay} {Regularization} (Jan
  2019), \url{http://arxiv.org/abs/1711.05101}, arXiv: 1711.05101

\bibitem{milickovic_ct_2000}
Milickovic, N., Baltast, D., Giannouli, S., Lahanas, M., Zamboglou, N.: {CT}
  imaging based digitally reconstructed radiographs and their application in
  brachytherapy. Physics in Medicine and Biology  \textbf{45}(10),  2787--2800
  (Oct 2000)

\bibitem{van_den_oord_neural_2017}
van~den Oord, A., Vinyals, O., kavukcuoglu, k.: Neural {Discrete}
  {Representation} {Learning}. In: Advances in {Neural} {Information}
  {Processing} {Systems}. vol.~30, pp. 6309--6318. Curran Associates, Inc.
  (2017)

\bibitem{ramesh_zero-shot_2021}
Ramesh, A., Pavlov, M., Goh, G., Gray, S., Voss, C., Radford, A., Chen, M.,
  Sutskever, I.: Zero-{Shot} {Text}-to-{Image} {Generation} (Feb 2021),
  \url{http://arxiv.org/abs/2102.12092}, arXiv:2102.12092 [cs]

\bibitem{schock_monoplanar_2022}
Schock, J., Lan, Y.C., Truhn, D., Kopaczka, M., Conrad, S., Nebelung, S.,
  Merhof, D.: Monoplanar {CT} {Reconstruction} with {GANs}. In: 2022 {Eleventh}
  {International} {Conference} on {Image} {Processing} {Theory}, {Tools} and
  {Applications} ({IPTA}). pp.~1--6 (Apr 2022), iSSN: 2154-512X

\bibitem{shen_patient-specific_2019}
Shen, L., Zhao, W., Xing, L.: Patient-specific reconstruction of volumetric
  computed tomography images from a single projection view via deep learning.
  Nature Biomedical Engineering  \textbf{3}(11),  880--888 (Nov 2019),
  \url{https://www.nature.com/articles/s41551-019-0466-4}, number: 11
  Publisher: Nature Publishing Group

\bibitem{shiode_2d3d_2021}
Shiode, R., Kabashima, M., Hiasa, Y., Oka, K., Murase, T., Sato, Y., Otake, Y.:
  {2D}–{3D} reconstruction of distal forearm bone from actual {X}-ray images
  of the wrist using convolutional neural networks. Scientific Reports
  \textbf{11}(1),  15249 (Jul 2021),
  \url{https://www.nature.com/articles/s41598-021-94634-2}, number: 1
  Publisher: Nature Publishing Group

\bibitem{vaswani_attention_2017}
Vaswani, A., Shazeer, N., Parmar, N., Uszkoreit, J., Jones, L., Gomez, A.N.,
  Kaiser, Å., Polosukhin, I.: Attention is {All} you {Need}. In: Advances in
  {Neural} {Information} {Processing} {Systems}. vol.~30. Curran Associates,
  Inc. (2017),
  \url{https://proceedings.neurips.cc/paper/2017/hash/3f5ee243547dee91fbd053c1c4a845aa-Abstract.html}

\bibitem{ying_x2ct-gan_2019-1}
Ying, X., Guo, H., Ma, K., Wu, J., Weng, Z., Zheng, Y.: {X2CT}-{GAN}:
  {Reconstructing} {CT} from {Biplanar} {X}-{Rays} with {Generative}
  {Adversarial} {Networks} (May 2019), \url{http://arxiv.org/abs/1905.06902},
  arXiv:1905.06902 [cs, eess]

\end{thebibliography}

%
%
%
%

\newpage
\section{Supplementary Material}
\renewcommand{\thefigure}{S\arabic{figure}}
\renewcommand{\thetable}{S\arabic{table}}
\setcounter{figure}{0}    
\setcounter{table}{0}   

\setlength{\tabcolsep}{0.5em}
{\renewcommand{\arraystretch}{1.4}
\begin{table}[]
\centering
\caption{Hyperparameters used to train the VQ-GAN and GPT models.}
\label{tab:hyperparams}
\resizebox{\columnwidth}{!}{%
\begin{tabular}{l|c|c|c|}
\cline{2-4}
\textbf{}                                             & \textbf{2D VQ-GAN} & \textbf{3D VQ-GAN} & \textbf{GPT Model} \\ \hline
\multicolumn{1}{|l|}{\textbf{Batch Size}}             & 50                 & 2                  & 12                 \\ \hline
\multicolumn{1}{|l|}{\textbf{Optimizer}}              & Adam \cite{kingma_adam_2015-1}              & Adam               & AdamW \cite{loshchilov_decoupled_2019}             \\ \hline
\multicolumn{1}{|l|}{\textbf{Learning rate}}          & 5e-6               & 5e-6               & 6e-4               \\ \hline
\multicolumn{1}{|l|}{\textbf{Scheduler}} &
  None &
  None &
  \begin{tabular}[c]{@{}c@{}}Cosine Annealing\\ (with warmup) \\ \cite{loshchilov_sgdr_2017} \end{tabular} \\ \hline
\multicolumn{1}{|l|}{\textbf{Epochs}}                 & 500                & 500                & 300                \\ \hline
\multicolumn{1}{|l|}{\textbf{No. Codebook Entries}}   & 8192               & 8192               & -                  \\ \hline
\multicolumn{1}{|l|}{\textbf{Encoder Depth}}          & 4                  & 4                  & -                  \\ \hline
\multicolumn{1}{|l|}{\textbf{\begin{tabular}[c]{@{}l@{}}Latent Dimension\\ (height, width, depth)\end{tabular}}} &
  16 x 16 x1 &
  16 x 16 x 16 &
  - \\ \hline
\multicolumn{1}{|l|}{\textbf{No. Transformer Blocks}} & -                  & -                  & 8                  \\ \hline
\multicolumn{1}{|l|}{\textbf{\begin{tabular}[c]{@{}l@{}}No. Heads for Multi-Head\\ Self-Attention\end{tabular}}} &
  - &
  - &
  8 \\ \hline
\multicolumn{1}{|l|}{\textbf{Embedding Dimension}}    & -                  & -                  & 512                \\ \hline
\end{tabular}%
}
\end{table}
}

\end{document}